\documentclass{jaa}

\usepackage{graphicx,txfonts,amsmath}
\usepackage[table]{xcolor}
\usepackage{float}

\newcommand{\HI}{\textup{H\,{\mdseries\textsc{i}}}~}
\newcommand{\HII}{\textup{H\,{\mdseries\textsc{ii}}}}

\newcommand{\kms}{{km~s$^{-1}$}}

\newcommand{\HaN}{{H$\alpha$+[N II]}\ }
\newcommand{\Ha}{{H$\alpha$}\ }

\def\kms{{km~s$^{-1}$}}

\def\arcmin{\hbox{$^\prime$}}
\def\arcsec{\hbox{$^{\prime\prime}$}}

\def\farcsec{\hbox{$.\!\!^{\prime\prime}$}}

\begin{document}\sloppy

\title{Dorado and its member galaxies II.  \\
	A {\tt UVIT} picture of the NGC 1533 substructure}


\author{Rampazzo R.\textsuperscript{1,2,*}, Mazzei P.\textsuperscript{2}, Marino A.\textsuperscript{2}, Bianchi L.\textsuperscript{3},  
	Ciroi S.\textsuperscript{4}, Held, E.V.\textsuperscript{2}, Iodice E.\textsuperscript{5}, Postma J.\textsuperscript{6},  Ryan-Weber, E.\textsuperscript{7}, 
	Spavone, M.\textsuperscript{5}, Uslenghi, M.\textsuperscript{8}}
\affilOne{\textsuperscript{1}INAF Osservatorio Astrofisico di Asiago, Via Osservatorio 8, 36012 Asiago, Italy.\\}
\affilTwo{\textsuperscript{2}INAF Osservatorio Astronomico di Padova, Vicolo dell'Osservatorio 5, 35122 Padova, Italy.\\}
\affilThree{\textsuperscript{3} Dept. of Physics \& Astronomy, The Johns Hopkins University, 3400 N. Charles St., Baltimore, MD 21218, USA\\}
\affilFour{\textsuperscript{4}  Dept. of Physics and Astronomy, University of Padova, Vicolo dell'Osservatorio 3, 35122 Padova, Italy\\}
\affilFive{\textsuperscript{5}  INAF-Osservatorio Astronomico di Capodimonte, Salita Moiariello 16, 80131 Napoli, Italy \\}
\affilSix{\textsuperscript{6} University of Calgary, 2500 University Drive NW, Calgary, Alberta, Canada\\}
\affilSeven{\textsuperscript{7} Centre for Astrophysics and Supercomputing, Swinburne University of Technology, Hawthorn, Victoria 3122, Australia\\}
\affilEight{\textsuperscript{8} INAF-IASF, Via A. Curti, 12, 20133 Milano (Italy)} 

\twocolumn[{

\maketitle

\corres{roberto.rampazzo@inaf.it}

\msinfo{31 October 2020}{}

\begin{abstract}
Dorado is a nearby (17.69 Mpc) strongly evolving galaxy group in the
Southern Hemisphere. We are investigating the star formation in this
group. This paper provides a FUV imaging of NGC 1533, IC 2038 and IC
2039, which form a substructure, south west of the Dorado group
barycentre. FUV CaF2-1  {\tt UVIT-}{\tt Astrosat} images enrich our
knowledge of the system provided by {\tt GALEX}. In conjunction with
deep optical wide-field, narrow-band \Ha\ and 21-cm radio images we
search for signatures of the  interaction mechanisms looking in  the FUV
morphologies and  derive the star formation rate. The shape of the FUV
luminosity profile suggests the presence of a disk in all three
galaxies. FUV emission is detected out to the optical size for IC 2038,
and in compact structures corresponding to \Ha\ and \HII\ bright
features in NGC 1533. A faint  FUV emission, without an optical
counterpart,  reminiscent of  the \HI\ structure that surrounds the
outskirts of NGC 1533 and extends up to IC 2038/2039, is revealed above
the local background noise.
\end{abstract}

}]


\doinum{12.3456/s78910-011-012-3}
\artcitid{\#\#\#\#}
\volnum{000}
\year{0000}
\pgrange{1--}
\setcounter{page}{1}
\lp{1}

\section{Introduction}

One of the breakthrough provided by the {\it Galaxy Evolution Explorer}
({\tt GALEX} hereafter) (Martin et al. 2005, Morrissey et al. 2007) is
the direct evidence of galaxy transformation in groups. UV - optical
colour magnitude diagrams (CMDs hereafter) highlighted an intermediate
region, the green valley, populated by transforming galaxies (see e.g.
Salim et al. 2007, Schawinski et al. 2007) located between the sequence
of red galaxies, mostly evolved early-type (Es+S0s=ETGs hereafter), and
the blue cloud composed of late-type (LTGs hereafter), star forming
galaxies. Ranking groups according to their blue vs. red sequence and
green valley galaxy populations, UV - optical CMDs contribute to
correlate the group structure, kinematics and dynamics to its members
evolutionary phase. Marino et al. (2010, 2013) investigated loose
groups, rich of LTGs, analogs of our Local Group.  This kind of groups
lack a well defined red sequence. At odd, less dispersed groups with an
increasing fraction of ETGs start to develop a galaxy population
inhabiting both the red sequence and the green valley. Very rich groups,
in an advanced stage of virialization, like NGC 5486 group, the third
richest association in the nearby universe after Virgo and Fornax, show
a well developed red sequence and a still rich green valley (Marino et
al. 2016, Mazzei et al 2014b, Mazzei et al. 2019 and references
therein). The enrichment of the red sequence and of the green valley
traces the transition from loose yet un-virialized groups to rich, more
compact and virialized  ones (see e.g. Rampazzo et al. 2018).

Driven by the gravitational force, that collapses groups and makes
galaxies to interact,  mechanisms leading to a galaxy morphological
metamorphosis can either quench or re-ignite the star formation (SF).
Mechanisms involved are of several types and  seem to depend on the
environment density (see e.g Boselli et al. 2006, Boselli et al. 2014).
Mergers can transform spirals into ellipticals and S0s (see e.g. Toomre
\& Toomre 1972, Barnes 2002, Mazzei et al. 2014a and references therein)
and quench SF by ejecting the interstellar medium via supernov\ae, AGN
or shock-driven winds (see e.g. Di Matteo et al. 2015 and references
therein). Ram-pressure, that may strip the gas reservoirs is supposed to
work mostly in rich environments (Boselli et al. 2014, Ramatsoku et al.
2019), but there are evidences (e.g. in \HI) that it also works for
groups (Bureau et al. 2002, Kantharia et al. 2005). There are examples
of mass--transfer between gas rich and gas poor companion galaxies, e.g
physical pairs composed of a LTG and an ETGs, that may re-fuel SF (see
e.g. Domingue et al. 2003, Keel et al. 2004, Chung et al. 2006, Plana et
al. 2017 ).

\begin{figure}
	\center
	{{\includegraphics[width=8.6cm]{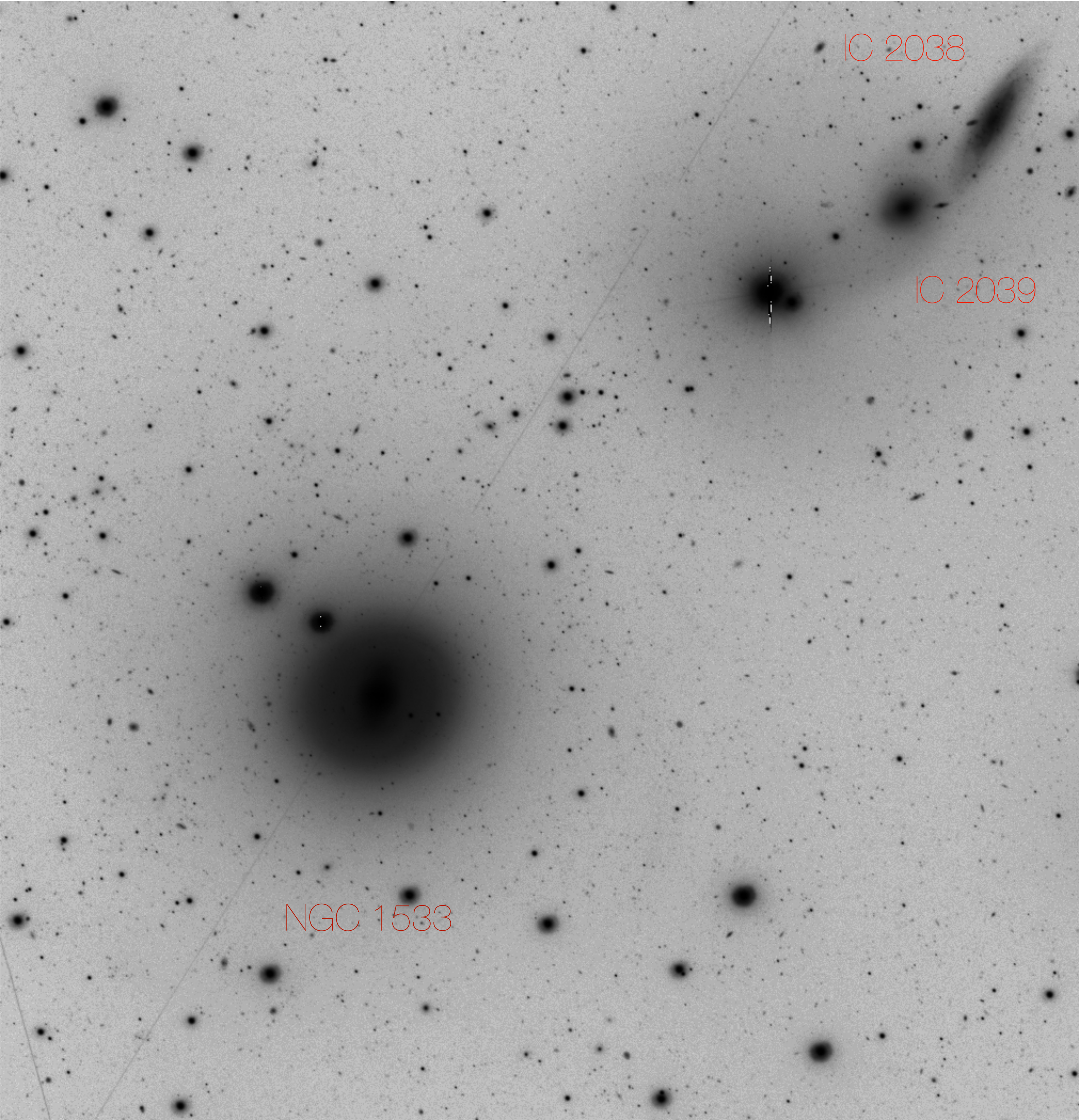}}
\caption{NGC 1533 sub-structure. VST deep image in the SDSS  $g$-band
of the IC 2038, IC 2039 (north west side) and NGC 1533 (south east side).
 The image size is 14\arcmin$\times$14.7\arcmin. 
{\tt Astrosat-UVIT} observations (see Table~\ref{UVIT-sources}), 
cover the entire sub-structure.
\label{Dorado-group}}}
\end{figure}

\medskip
In this paper, we focus on the nearby   Dorado group in the Southern
Hemisphere (RA=64.3718 [deg], Dec=-55.7811 [deg]) as defined by 
(Kourkchi \& Tully (2017; see also Firth et al. 2006 and references
therein). Throughout the paper we adopt 17.69 Mpc as the distance of all
Dorado candidate members. Dorado CMD is described by Cattapan et al.
(2019, their Figure 1). The red sequence of the group includes several
ETGs. A set of intermediate luminosity member galaxies is still crossing
the green valley. Basically only NGC 1566, a bright grand design spiral,
is still located in the blue cloud. With respect to  evolved groups
(e.g. NGC 5486, see Marino et al. 2016) Dorado seems to be in an earlier
and active evolutionary phase. Several indications support this view
like the group clumpy structure. A compact group, SGC~0414-5559 (Iovino
2002) is located at the barycentre of the group as defined by Kourkchi
\& Tully (2017). The compact group is dominated by two ETGs, NGC 1549 an
E and NGC 1553 an S0, both showing a wide shell structure. The Dorado
group members morphology shows indeed the nearly ubiquitous presence of
interaction signatures such as shells, asymmetries, tails. (see e.g.
Malin \& Carter 1988, Cattapan et al. 2019). Star forming rings have
been also revealed in  several ETGs (Rampazzo et al. 2020). Dorado is an
atomic gas rich group. Nearly half of the entire \HI\ content of the
Dorado group, 3.5$\times$10$^{10}$ M$_\odot$, is located in the spiral
member NGC 1566, although \HI\ has been also detected  in several other
members, independently from their morphological classification (Kilborn
et al. 2005, 2009). Rampazzo et al. (2020), presented their \HaN\ study
aiming at investigating the star formation rate (SFR hereafter) of the
Dorado backbone galaxies. They found that SFR in LTGs is fading while in
ETGs is not yet shut-down. Rampazzo et al. (2020) suggested that
mechanisms such as gas stripping and gas accretion, through
galaxy-galaxy interaction, seem relevant in the evolutionary phase of
Dorado.

\begin{center}
	\begin{table*}[ht]%
		\centering
		\caption{Relevant properties of members of Dorado 
			in the UVIT FoV \label{UVIT-sources} from the literature}%
		\tabcolsep=0pt%
		\begin{tabular*}{40pc}{@{\extracolsep\fill}lcccccc@{\extracolsep\fill}}
			\hline
			\textbf{Field} &  \textbf{ID} &  \textbf{RA} & \textbf{Dec.}  & \textbf{FUV}  & \textbf{V$_{hel}$}   & 
			\textbf{Morpho.}  \\
			\textbf{centre}          &  \textbf{source}    &  \textbf{(J2000)}     & \textbf{(J2000)}            & \textbf{[mag]}  & \textbf{km s$^{-1}$} & \textbf{Type} \\
			\hline
			A       & {\bf IC 2038}  &  04 08 53.76  &  -55 59 22.2  & 17.25$\pm$0.02* &     712  & 7.0\\
			& {\bf IC 2039}  &  04 09 02.37 &   -56 00 42.1 & 19.98$\pm$0.07* &     817  & -3.1\\
			\hline
			B       &  {\bf NGC 1533}  &  04 09 51.84  &   -56 07 06.6  &  16.90$\pm$0.02$^1$  &  764     & -2.5\\
			\hline 
		\end{tabular*}
		\tablenotes
		{Column~1: {\tt UVIT} field. Column~2: source identification. 
			 Columns 3 and 4 provides Right Ascension and Declination of the sources. 
			  Column 5  lists the FUV
			total integrated magnitudes from {\tt GALEX} Archive (indicated
			with an *) and reported in {\tt NED} (indicated with $^1$).
			Columns 6 and  7 report the heliocentric radial velocity and
			the galaxy morphological Type, respectively, from Kourkchi \& Tully (2017).
		}
	\end{table*}
\end{center}

This work complements the Rampazzo et al. (2020) study, using FUV CaF2-1
(1300-1800 \AA) broad filter (similar to {\tt GALEX} FUV; see Tandon et
al. 2017) observations obtained with {\tt Astrosat-UVIT} of the Dorado
members. The {\tt UVIT} targets partly cover the galaxy set observed  in
\HaN\ by Rampazzo et al. (2020). In this paper we investigate the
sub-structure, SW of the Dorado barycentre, formed by NGC 1533, IC 2038
and IC 2039. Our goal is to analyze the FUV morphological structure of
galaxies in this sub-structure and the relation between \Ha\ regions and
the FUV emission. \Ha\ emission has been found not only in the Scd
galaxy IC 2038 but also in the E-S0 galaxy NGC 1533 (Rampazzo et al.
2020).

The paper plan is the following.  In Section~2 we summarize our
knowledge about the NGC 1533 sub-structure.  {\tt UVIT} observations and
the reduction are presented in Section~3. In Section~4 the
photometric analysis and the comparison with {\tt GALEX} observations
are presented.  Results and the discussion in Section~5 are given in
Section~6, respectively.

\section{NGC 1533 substructure and the Dorado group}
\label{NGC1533_literature}

According to Kourkchi \& Tully (2017),  Dorado has an average redshift
of 1230$\pm$89 km~s$^{-1}$ and a velocity dispersion of 242 km~s$^{-1}$.
NGC 1533 and the pair IC 2038/39 form a sub-structure of the group with
an average redshift of 764$\pm$43 km~s$^{-1}$ (see
Table~\ref{UVIT-sources}). Figure~\ref{Dorado-group} shows a deep
image of the NGC 1533 sub-structure in the SDSS $g$-band 
($\mu_g\approx30$ mag~arcsec$^{-2}$) obtained within the {\it VST 
Early-type GAlaxy Survey (VEGAS\footnote{Visit the website
http://www.na.astro.it/vegas/VEGAS/Welcome.html}) (see e.g. Capaccioli
et al. 2015).} This substructure is located at the south west periphery
of Dorado group barycentre formed by the SCG 0414-5559 compact group,
well separated, both in radial velocity and in projection, from other
member candidates.

NGC 1533 has been investigated with {\tt GALEX} by Marino et al.
(2011a,2011b). Rampazzo et al. (2017) investigated this galaxy with {\tt
Swift}. All these studies evidenced an outer FUV bright ring. In
correspondence to the FUV ring, Rampazzo et al. (2020) found some \Ha\
complexes. Moreover they found that while the SFR is enhanced in NGC
1533 it is depressed in IC 2038, if compared with the general sample of
ETGs (Gavazzi 2018) and of Spirals (James et al. 2004), respectively.
Rampazzo et al. (2020) found that in IC 2038 \HII\ regions are
distributed along the galaxy body, which appears slightly elongated
towards the companion galaxy, IC 2039, in the SE direction.

More recently, deep optical observations, in $g$ and $r$ bands, of NGC
1533 and IC~2038/39 have been analysed by Cattapan et al. (2019) using
the VLT Survey Telescope (VST) at the European Southern Observatory,
Chile. They evidenced a large disk around NGC 1533 and several tails
suggesting that NGC 1533 and the nearby pair are evolving together
(Cattapan et al. 2019).

The NGC 1533 Dorado substructure has been mapped in \HI\ by Ryan-Weber
et al. (2004) (see also Kilborn et al. 2005,2009) showing and extended
\HI\ structure that extends from 1533 up to IC 2038/39 pair. Werk et al.
(2008,2010) reported that there are some star forming regions well
outside NGC 1533 with the same radial velocity as the \HI\ gas revealed
by Ryan-Weber et al. (2004; see their Figure 7 for J0409-56). The
narrow-band \HaN\ study by Rampazzo et al. (2020) detected three \Ha\
regions in correspondence of the  regions detected by Werk et al. (2010)
confirming that such outer \HII\ regions belong to NGC 1533
sub-structure.

In this context, the study of the NGC 1533 sub-structure is relevant
both for the general understanding of the evolution of gas-rich merging
events and for the study of local SF mechanisms.

\begin{center}
	\begin{table*}
		\centering
		\caption{{\tt UVIT} observations \label{UVIT-observations}}%
		\tabcolsep=0pt%
		\begin{tabular*}{30pc}{@{\extracolsep\fill}lccccc@{\extracolsep\fill}}
			\hline
			\textbf{Field} &  \textbf{Obs ID} &  \textbf{Observing date} & \textbf{Exp. Time}  & \textbf{Target}  \\
			\textbf{ID }          &  \textbf{}    &  \textbf{}     & \textbf{[s]}            &   \textbf{ID }  \\
			\hline
			A  & A07\_010T04\_9000003220  & October 5, 2019 &  6481.278  & IC 2038  \\
			B  & A07\_010T05\_9000003236  & October 16, 2019 &  6628.734  & NGC 1533 \\
			\hline
		\end{tabular*}
		\tablenotes
{Field identification in col. 2 refers to the proposal A07$\_$010 (PI. R.
Rampazzo). Col. 3 and col. 4  report the observing date and the total
reduced exposure time. In col. 5 is quoted the central target. The
nominal zero point magnitude of the FUV channel and the physical plate
scale are 18.08 mag and 3\farcsec33 respectively,  as reported  in
Tandon et al. (2017).}
	\end{table*}
\end{center}
\medskip

\section{Observations and reduction}
\label{Observations}

{\tt Astrosat} is a multi-wavelength satellite that has been launched by
the Indian Space Research Organization on September 28, 2015. The
ultraviolet-optical  telescope on board is the Ultra-Violet Imaging
Telescope facility {\tt UVIT} (Tandon et al. 2017). It is composed of
two Ritchey-Chretien telescopes with  37.5 cm aperture,  a circular
field of view of 28' diameter, observing simultaneously one in FUV
(1300-1800 \AA) and the other  both in NUV (2000-3000 \AA) and optical
band, VIS (3200-5500 \AA), by means of a beam-splitter directing NUV and
VIS to individual cameras.

Since the NUV channel was not operative during our runs, observations
have been performed with the FUV channel only. We used  the full field
of view, in photon counting mode with the Filter F148W CaF2
($\lambda_{mean}$=1481, $\Delta \lambda$=500\AA).  Photons are counted
on a planar CMOS array at approximately 28~Hz and stacked on the ground
with shift and add algorithms (see for details Kumar et al. 2012, Postma
et al. 2011, Postma \& Leahy 2017, Tandon et al. 2017) with the
astrometric world coordinate solution solved automatically by a
trigonometric algorithm (Postma \& Leahy 2020).

Table~\ref{UVIT-observations} reports the relevant characteristics of
our {\tt UVIT} observations. {\tt Astrosat-UVIT} observations are the
result of the proposal A07\_010 (PI R. Rampazzo) and cover the south
west part of Dorado, in particular {\tt Astrosat-UVIT} fields  contain
IC 2038 and IC 2039 (Field A) and NGC 1533 (Field B).

\section{Data Analysis and comparison with the literature}
\label{Data-Analysis}

In this section we present the data analysis performed and the
comparison with the current literature mainly from {\tt GALEX study}. 
Original images with 0\farcsec416 per sub-pixel have been rebinned 
4$\times$4  providing a final scale of 1\farcsec664 px$^{-1}$.

\subsection{Integrated FUV magnitudes} \label{A1}

\begin{figure}
	\center
	{\includegraphics[width=8cm]{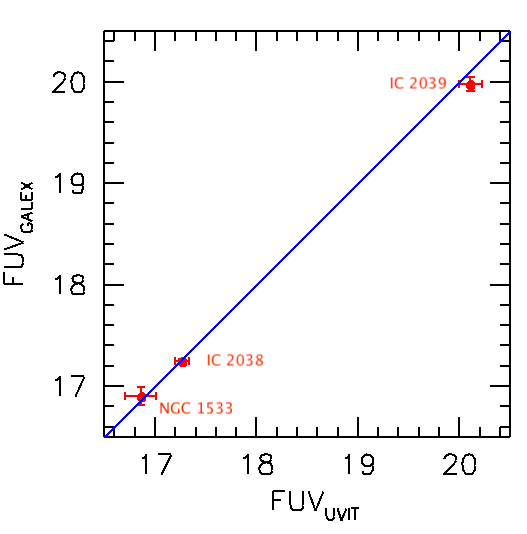}}
	\caption{Comparison between {\tt UVIT} FUV CaF2-1 integrated magnitudes of 
Dorado members with those from {\tt GALEX} available in the literature.
Magnitudes in the plot are not corrected for  galactic extinction.}
	\label{comparison_mag}
\end{figure}

Column 3 of Table~\ref{UVIT-results} provides the FUCaF2-1, {\tt UVIT}
integrated magnitudes. In Figure~\ref{comparison_mag} we compared 
our results with integrated magnitudes reported in Table~\ref{UVIT-sources} 
from {\tt GALEX} observations. The figure shows the good agreement
between our values and those in the current literature.

\subsection{Surface photometry}

The surface photometry is obtained using the {\tt IRAF} task {\tt
ELLIPSE} (Jedrzejewski 1987).  In obtaining the surface brightness
profile, {\tt ELLIPSE} accounts for the geometrical information
contained in the isophotes and allows the variation of the ellipticity,
$\epsilon=(1-b/a)$ and of the position angle, $(PA)$, along the ellipse
major axis ($a$). IN addition, {\tt ELLIPSE} provides measure of  the
isophotal shape parameter, the so-called $a_4$ parameter from the fourth
cosine component of the Fourier analysis of the fitted ellipse, allowing
to distinguish between boxy ($a_4<0$) and discy ($a_4>$0) isophotes
(Bender et al. 1988). This approach, widely used with optical images
of ETGs,  has been adopted to investigate {\tt GALEX} FUV data by Jeong
et al (2009) and Marino et al. (2011b). FUV geometric profiles
have not been provided by the above papers.

In fitting isophotes  we allowed $\epsilon$, and PA to variate with
the galacto-centric distance, in order to obtain a good description of
the FUV luminosity distribution. Our FUV images either show irregular
peculiar features (NGC 1533), clumpy and spiral features (IC 2038) or
have a low signal-to-noise (IC 2039). This is the reason for which we
provide in Table~\ref{UVIT-results} only the average ellipticity,
$\langle \epsilon \rangle$ and position angle, $\langle P.A. \rangle$,
(colums 4 and 5 respectively) of the galaxy.  In presence of irregular,
peculiar features, sudden variations in both $\epsilon$ and $PA$ are
expected and, in particular, the isophotal shape parameter, $a_4$,
looses its physical meaning.

\begin{figure}
	\center
	{\includegraphics[width=8.8cm]{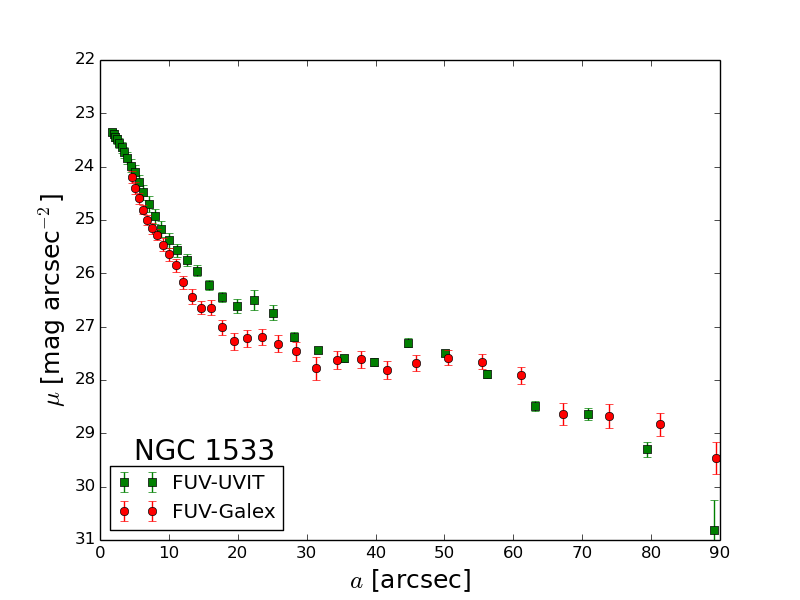}}
	\caption{Comparison of the present FUV luminosity profile of NGC 1533 with 
Marino et al. (2011b) from {\tt GALEX} pointed observations.   
		\label{NGC1533-comparison}}
\end{figure}

The pair IC~2038/IC~2039 has not been previously investigated with
{\tt GALEX}.  Marino et al. (2011b) obtained the surface brightness
profiles NGC 1533.  Figure~\ref{NGC1533-comparison} shows the comparison
between the Marino et al. (2011b) and our luminosity profile. The
luminosity profiles compare quite well outside 30\arcsec, suggesting
that there is a very small, if any, zero point effect, while the central
region differs significantly. The reason of such a difference in the
central part is not entirely clear, although in FUV  the {\tt GALEX} PSF
FWHM is 4.2\arcsec with respect to 1.5\arcsec of {\tt UVIT} (see
Morrissey et al. 2007 and Tandom et al. 2017). Substructures on spatial
scales (projected on the sky) $\lesssim0.4$ kpc are washed out in the
{\tt GALEX} profile, but are picked up in the {\tt UVIT} data. In
particular, the FUV knot that peaks at $\approx$25\arcsec (indicated as
A in Figure~\ref{Tricromia_NGC1533}) is clearly smoothed out in Marino
et al. (2011b), suggesting a significant role played by the PSF in
driving the {\tt GALEX} surface photometry.

Figure~\ref{Tricromia_IC2038} and Figure~\ref{Tricromia_NGC1533}
show the FUV image and the luminosity profile of the pair members
IC~2038/39 and of NGC 1533, respectively. The bottom panel of these
figures shows a color composite RGB image of the galaxies obtained from
the present FUV image, the continuum and \HaN\ images from Rampazzo et
al. (2020).

\subsection{Luminosity profile fitting}
\medskip
We fit the shape of the FUV luminosity profile with a S\'ersic law
(S\'ersic 1963).  Even a crude representation of the light profile, as a
simple S\'ersic law fit, provides useful information, and sometimes is
the only decomposition that can be compared with the literature. The
S\'ersic law, $\mu\propto r^{1/n}$ where $\mu$ is the surface brightness
and $r$ the radius, is a generalization of the de Vaucouleurs (1953)
$r^{1/4}$ and of the Freeman (1970) exponential laws. The variation of
the S\'ersic index, $n$, describes the manifold of the shapes of 
luminosity profiles  of ETGs. The watershed can be  considered the value
$n=4$ representing a `classic' elliptical. A classic exponential disc
(Freeman 1970) in S0s has an index $n=1$.

FUV luminosity profiles of ETGs may reach large values of $n$ (see e.g.
Marino et al. 2011b). Rampazzo et al. (2017) suggested that in
luminosity profiles with $n<3$ the presence of a disc starts to emerge.

\begin{figure*}
	\center
	{\includegraphics[width=5.5cm]{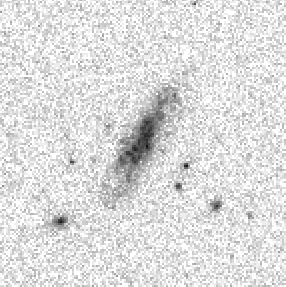}}
	{\includegraphics[width=5.9cm]{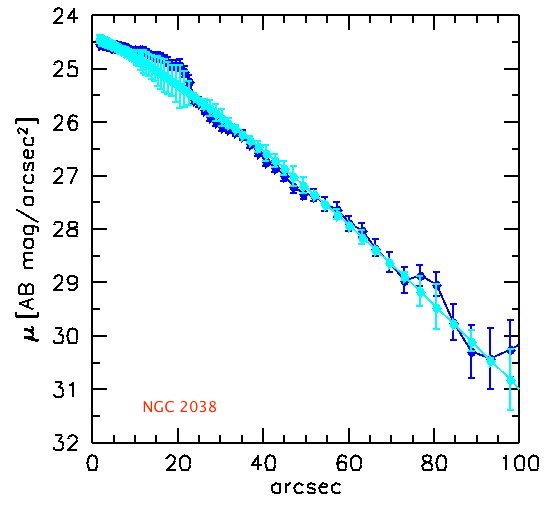}}
	{\includegraphics[width=5.9cm]{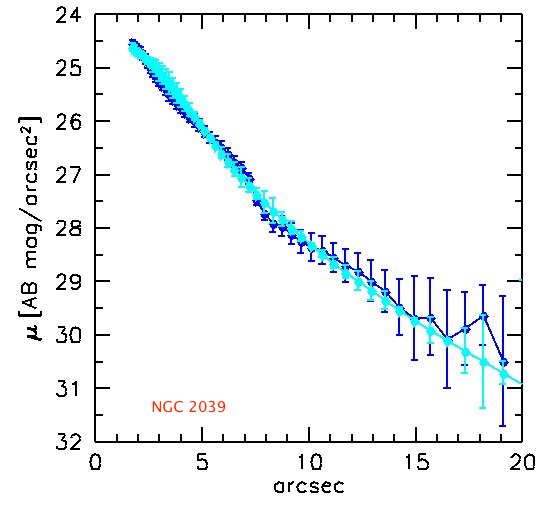}}
	{\includegraphics[width=12.5cm]{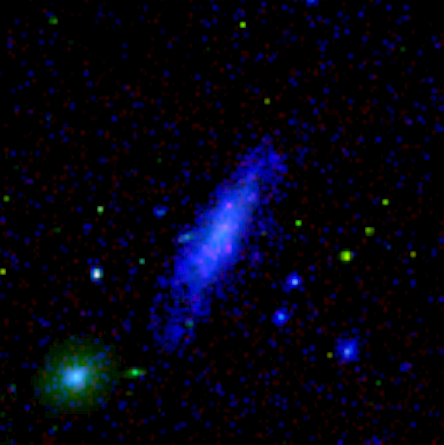}}
	\caption{({\it Top: left panel)} FUV image of IC 2038 (Sbc) and of  IC 2039
(E), its physical companion, in the south east. The image size is
4\arcmin$\times$4\arcmin. North on the top, east to the left. ({\it Top:
mid and right panels}) FUV surface brightness profile of IC 2038 and of
IC 2039 (blue dots). Single S\'ersic law fit of the luminosity profiles,
discussed in Section~\ref{Results} are also shown as cyan dots. The
values of the S\'ersic index are  $n=0.81\pm0.09$ and $n=1.45\pm0.18$,
for IC 2038 and IC 2039, respectively. ({\it Bottom panel})
Color-composite RGB image using the red and the green channels for \Ha\
and  the nearby continuum (Rampazzo et al. 2020) and the blue channel
for the FUV image. Both the \Ha\ and the continuum PSF have been
re-scaled to the PSF of the FUV image.
		\label{Tricromia_IC2038}}
\end{figure*}

\begin{figure*}
	\center
	{\includegraphics[width=5.5cm]{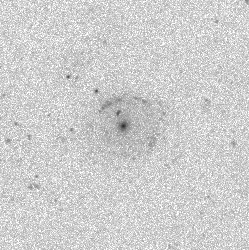}}
{\includegraphics[width=6.5cm]{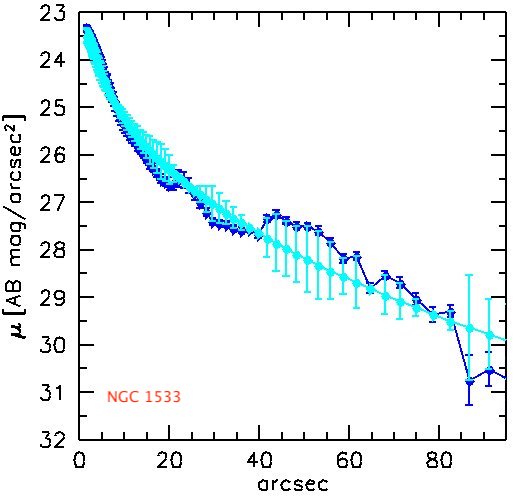}}
	{\includegraphics[width=12.5cm]{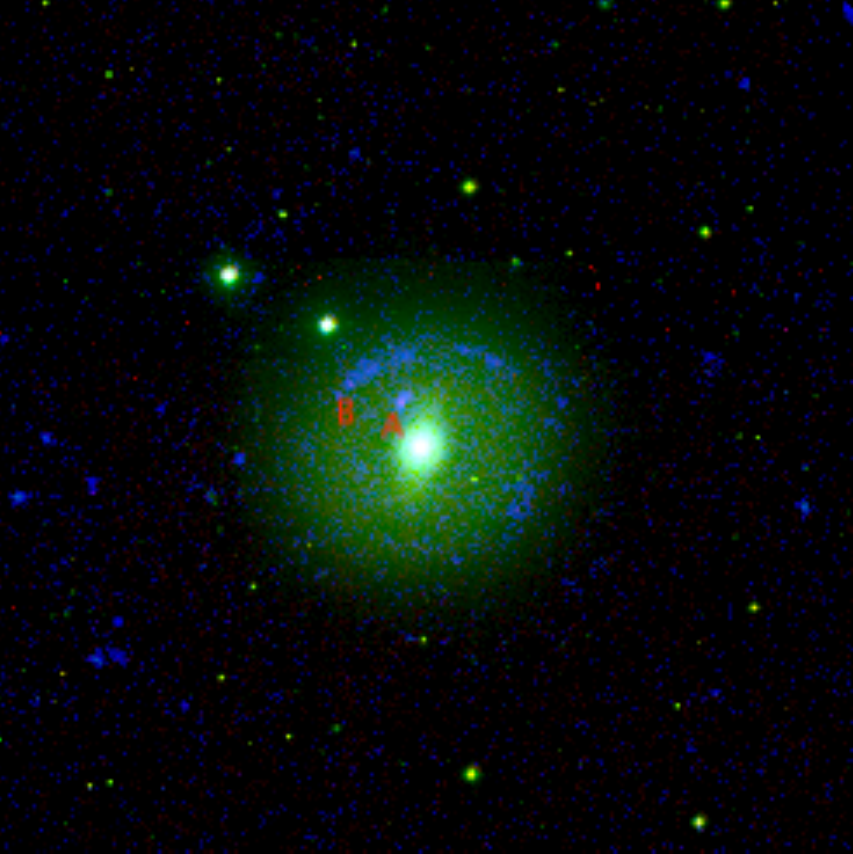}}
	\caption{({\it Top: left panel)}  FUV image of NGC 1533. The image size is
7\arcmin$\times$7\arcmin. North on the top, east to the left. ({\it Top:
right panel}) Surface brightness profile of NGC 1533 (blue dots) and
single S\'ersic law fit (cyan dots). The value of the S\'ersic index is 
$n=2.64\pm0.06$. ({\it Bottom panel}) As in
Figure~\ref{Tricromia_IC2038} for NGC 1533. Labels A and B indicate
\HII\ regions in Rampazzo et al. (2020).
		\label{Tricromia_NGC1533}}
\end{figure*}

The S\'ersic law fit is shown  in  the right panel/s of
Figures~\ref{Tricromia_IC2038} and of Figure~\ref{Tricromia_NGC1533} for
IC2038/IC 2039 and NGC 1533, respectively, superposed to the luminosity
profile. The fit is extended to the entire profile without masking FUV
bright sub-structures such as  the ring (B) and the knot (A)  in NGC
1533. The S\'ersic fit accounts for the {\tt UVIT}-PSF so it is not
necessary to avoid the inner part of the luminosity profile certainly
`contaminated' by the instrument PSF.

\begin{table*}[t]%
	\centering
	\caption{Relevant FUV parameters of galaxies in the NGC~1553 sub-structure}
	\label{UVIT-results}%
	\begin{tabular*}{40pc}{@{\extracolsep\fill}lccccccc@{\extracolsep\fill}}%
		\hline
		\textbf{Field} &  \textbf{ID}     & \textbf{FUV}   & \textbf{$\langle \epsilon \rangle$} & \textbf{$\langle PA \rangle$}  & \textbf{n} & \textbf{L$_{FUV}$} & \textbf{SFR}\\
		\textbf{ }     &  \textbf{source} & [mag] &            & [deg]&    & [erg~s$^{-1}$~Hz$^{-1}$] &[M$_\odot$ yr$^{-1}$]\\
		\hline
		A  &  {\bf IC 2038} & 17.19$\pm$0.07  & 0.72$\pm$0.02  & 154.4$\pm$4.6 & 0.81$\pm$0.09    & 1.81$\times$10$^{27}$   & 0.025$\pm$0.002  \\
		&  {\bf IC 2039} & 20.03$\pm$0.11  & 0.19$\pm$0.05  & 104.1$\pm$8.7 & 1.45$\pm$0.18   & 1.33$\times$10$^{25}$&0.002$\pm$0.0002   \\
		\hline
		B  & {\bf NGC 1533} & 16.74$\pm$0.15  & \dots   & \dots  & 2.64$\pm$0.06 &2.74$\times$10$^{26}$     &  0.038$\pm$0.006  \\
		\hline \hline
	\end{tabular*}
	\tablenotes{Column 1 {\tt UVIT} field. Column 2 Source identification; column 3  
FUV integrated magnitude corrected for galactic extinction accounting
for  A$_{FUV}$=7.9$\times$E(B-V); E(B-V) is 0.01 mag  both for IC2038
and IC 2039, 0.015 mag for NGC1533 from NED. Luminosities are computed
accounting for  the same distance, 17.69 Mpc, for all our targets. In
columns 4 and 5 are provided the average ellipticity and position
angles: for NGC 1533 see Section~\ref{Results} In column 6 is reported
the S\'ersic index from the best-fit of the entire profile. In columns 7
and 8 are reported the FUV luminosity and the SFR computed according to
the Lee et al. (2009) recipe provide in Section~\ref{SF_FUV}. 
	}
\end{table*}

\section{Results}
\label{Results}

In the following sections we discuss the shape of the surface brightness profile
and the morphology of the FUV emission.

\subsection{Members surface photometry}
\label{surface_brightness} 
\medskip
\noindent
\underbar{IC 2038}

In Figure~\ref{Tricromia_IC2038} top panels the FUV image and the
luminosity profiles of IC 2038 and IC 2039 are shown.

IC 2038, classified as Sbc in optical by {\tt HyperLeda}, does not show
a bulge (see e.g. Rampazzo et al. 2020 their Figure~3). The FUV emission
appears clumpy in the centre of the galaxy, where \HII\ regions are
detected by Rampazzo et al. (2020), while  an arm-like structure appears
out the both sides of the galaxy body.

The FUV surface brightness profile is flat in the central
$\approx$20\arcsec\. The best fit to the entire profile has a S\'ersic
index of $n=0.81\pm0.09$ suggesting the galaxy is dominated by the disk.
 In the central $\approx$25\arcsec\ there is an excess of luminosity
with respect to the S\'ersic law best fit,  in correspondence to FUV
clumps.

The FUV emission extends out to the optical size of the galaxy as shown
by Figure~\ref{Tricromia_IC2038} and Figure~\ref{Dorado-group}: there is
no evidence of a XUV disk (Thilker 2008).

The average ellipticity, $\langle\epsilon\rangle=0.72\pm0.02$ and the
average position angle $\langle PA \rangle=154.4^\circ\pm4.6^\circ$,
compare well with values from optical bands, 0.74$\pm$0.04 and
155.2$^\circ$ respectively, from the {\tt HyperLeda} catalog.

\medskip
\noindent
\underbar{IC 2039}

In FUV IC 2039 is  the faintest galaxy of the sub-structure 
(Figure~\ref{comparison_mag}). We best fit  luminosity profile of IC
2039 (top right panel of Figure~\ref{Tricromia_IC2038}) with a single
S\'ersic law with an index $n=1.45\pm0.18$ indicating that it is
basically composed of a disc.

The average ellipticity is $\langle\epsilon\rangle=0.19\pm0.05$ and the
position angle $\langle PA \rangle=104.1^\circ\pm8.7^\circ$ to compare
with optical values of 0.15$\pm$0.07 and 124.3$^\circ$ from {\tt
HyperLeda}.

\medskip
\noindent
\underbar{NGC 1533}

The galaxy, classified as E-S0 ({\tt HyperLeda}) is known to have an
outer ring and a inner bar, so  Comer\'on et al. (2014)  classified it
as (RL)SB0$^0$. However, in FUV this galaxy shows no signature of the
bar. The presence of a FUV-bright outer ring was evidenced using {\tt
GALEX} by Marino et al. (2011b) both in NUV and FUV. {\tt UVIT} clearly
reveals a bright FUV spot at $\approx$25\arcsec\ and the ring in the
range 39\arcsec -- 81\arcsec, in agreement with Marino et al. 2011
(their Table 1). Their presence causes a sudden jump of both
$\epsilon$ and PA. {\tt ELLIPSE} provides $\epsilon=0.56\pm0.04$ and
PA$=63^\circ$ within 39\arcsec\ and $\epsilon\approx$0.17$\pm0.06$ and
PA=$164^\circ$ outside the region of the ring up galaxy outskirts. The
value of the ellipticity and PA provided by {\tt HyperLeda} from optical
bands are 0.36$\pm$0.09 and 141.5$^\circ$.

The surface brightness profile (Figure~\ref{Tricromia_NGC1533} top right
panel) is best-fitted by a S\'ersic law with index $n= 2.64\pm0.06$
suggesting the presence of a disc. This value agrees with the UV
S\'ersic index  of $n= 2.76\pm0.10$ obtained from the luminosity profile
in the {\tt Swift} W2 filter ($\lambda_c$=2030 \AA) by Rampazzo et al
(2017).

\begin{figure*}
	\center
	{\includegraphics[width=8cm]{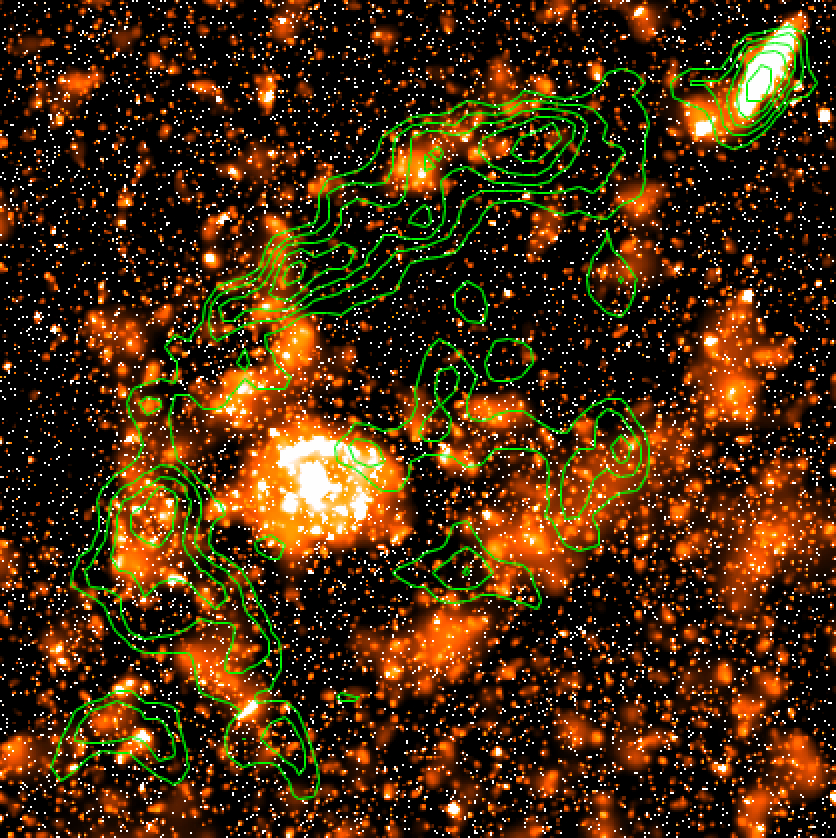}}
	{\includegraphics[width=8cm]{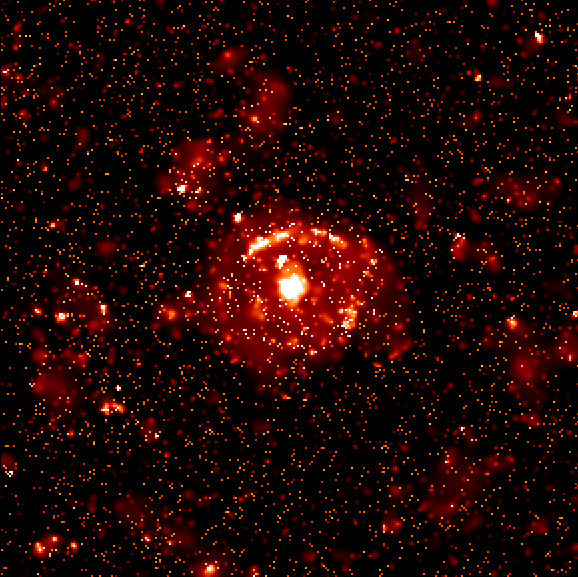}}
	\caption{ {\it (Left panel)} The \HI\ contour levels from Ryan-Weber  et al.
(2004) at column densities 2.5$\times$10$^{20}$, 2.8$\times$10$^{20}$,
3.1$\times$10$^{20}$, 3.5$\times$10$^{20}$, 3.9$\times$10$^{20}$ and
4.2$\times$10$^{20}$ atoms~cm$^{-2}$ are superposed 	to the {\tt
UVIT} FUV image of the sub-structure after an adaptive smoothing ({\tt
ASMOOTH}) with a minimum S/N $\tau_{min}=1.5$ (Ebeling et al. 2006) has
been applied. The image size corresponding to Figure~\ref{Dorado-group}.
{\it (Right panel)} Zoom-in on NGC 1533 (field of View
7\arcmin$\times$7\arcmin) showing the complex structure of the ring and
an arm/like structure on the west side ({\tt ASMOOTH} with a minimum S/N
$\tau_{min}=2$). }
	\label{NGC1533_HI_UV}
\end{figure*}

\subsection{FUV regions outside galaxies main body}
\label{FUV_outside}

To enhance the signal-to-noide ratio (S/N) in the galaxy outskirts and
bring out any possible faint structures in the UV emission, we adopted
the procedure outlined by Ebeling et al. (2006), called {\tt ASMOOTH}.
The only parameter required by by the procedure is the desired minimal
S/N, $\tau_{min}$. For each individual pixel, the algorithm increases
the smoothing scale until the S/N within the kernel reaches a specified
input value. {\tt ASMOOTH} suppresses very efficiently the noise while
the signal, locally significant at the selected S/N level, is preserved
on all scales. In the right and left panels of
Figure~\ref{NGC1533_HI_UV}, the FUV image has been treated with {\tt
ASMOOTH} selecting a S/N above the back-ground of $\tau_{min}$=1.5 and
$\tau_{min}$=2.0, respectively. The possible physical causes of the
features that emerged using this procedure are discussed below.

Cattapan et al. (2019; see their Figure~7) superposed the emission of
NGC 1533 in  the W2 filter derived by Rampazzo et al. (2017) from {\tt
Swift-UVOT} observations to their wide field, deep g-band image obtained
at VST (see Figure~\ref{Dorado-group}).  They found that the FUV ring of
NGC 1533 is superposed to  spiral-like residuals obtained after a model 
of the optical luminosity profile has been subtracted. The right  panel
of Figure~\ref{NGC1533_HI_UV}, the FUV image treated with {\tt ASMOOTH}
shows an arm-like structure is emerging also in FUV outside the ring in
the west side.

In the left panel of Figure~\ref{NGC1533_HI_UV} we overplot the \HI\
emission isophotes from Ryan-Weber et al. (2004). While the center of
NGC 1533, in particular the ring, is devoid of \HI\, the large scale
structure of \HI\  seems to corresponds to the faint extended FUV
emission derived using {\tt ASMOOTH}. The ATCA \HI\ observations by
Ryan-Weber et al. (2004) extended over an area at least
$\approx$5\arcmin\ to west (partly covered by the
Figure~\ref{NGC1533_HI_UV}). They do not detect any \HI\ to the west
within their detection limit. Some FUV emission regions above
$\tau_{min}$=1.5 are still visible in that area. The presence of this
FUV emission would be consistent with the N-body SPH  numerical
simulations by Ryan-Weber et al. (2003) describing the \HI\ ring around
NGC 1533 as the remnant of a tidally destroyed galaxy (see Section~6.2).

\begin{figure*}
	\center
{\includegraphics[width=10cm]{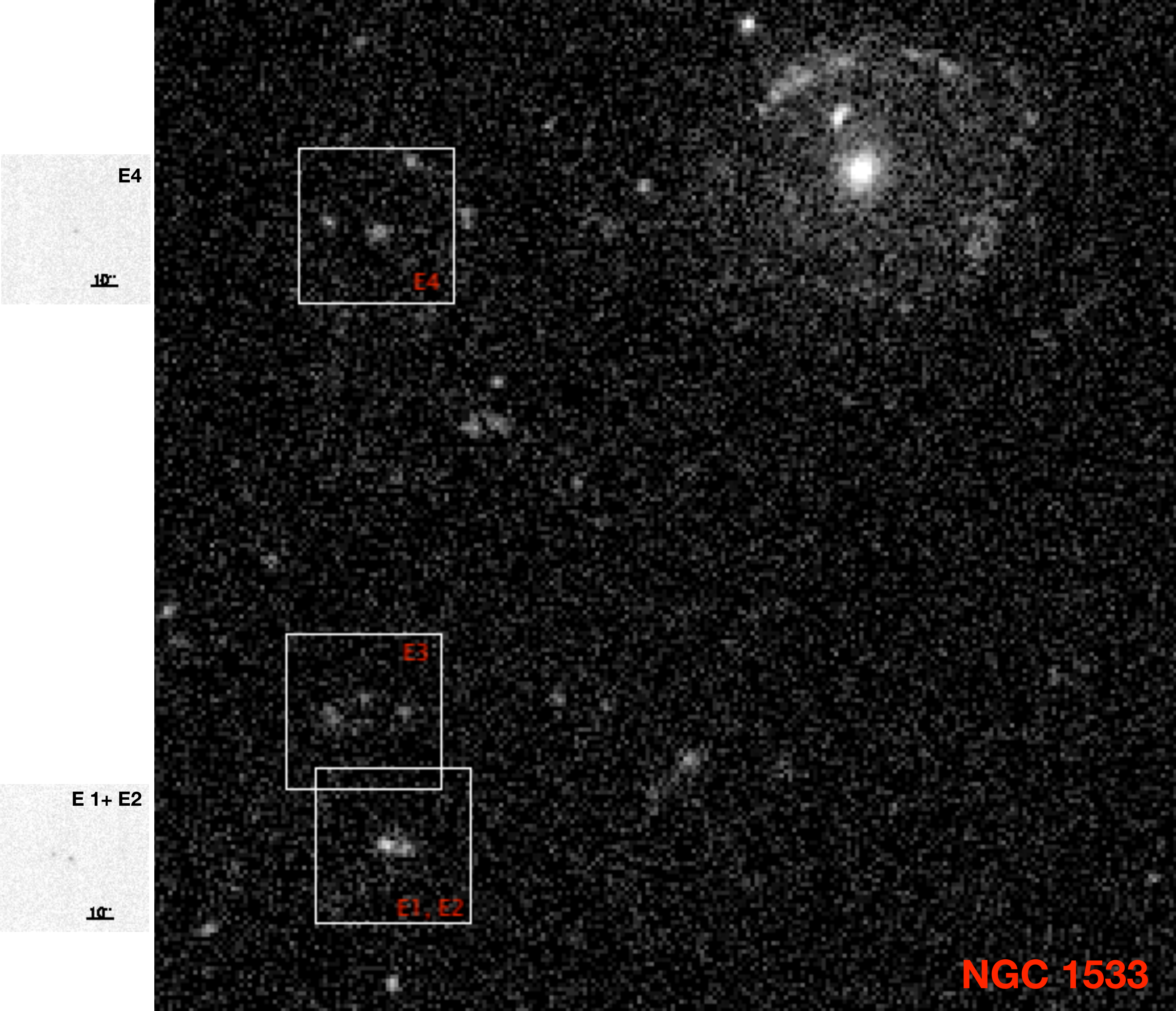}}
	\caption{{\bf The  south-east region of NGC 1533 observed in FUV by {\tt UVIT}.
Three boxes of 60\arcsec$\times$60\arcsec in the FUV image enclose the
regions identified as J0409-56 E1 + E2, E3, E4 by Werk et al. (2010) on
{\tt GALEX} in which they found \Ha\ sources. The two areas, showed as
sub-panels on the right of the FUV image,} mark the position of \HII\
regions, at the same redshift of the galaxy, detected by Rampazzo et al.
(2020) via \HaN\ observations. The centres of the regions in Rampazzo et
al. (2020) correspond to the following coordinates 04 10 13.5 -56 11 36
(J2000) and 04 10 10.66 -56 07 27.99 (J2000) and overlap with region
E1+E2 and E4 in Werk et al. (2010), respectively.
\label{NGC1533_Ha_outer}}
\end{figure*}

\medskip
Figure~\ref{NGC1533_Ha_outer}  shows the outer FUV emission regions of
NGC 1533, investigated  by Werk et al. (2008, 2010) in the first
systematic search for outlying \HII\ regions, as part of a sample of 96
emission-line point sources (referred to as ELdots–emission-line dots)
derived from the NOAO Survey for Ionization in Neutral Gas Galaxies
(SINGG). Such regions, selected from {\tt GALEX} FUV, are indicated in
the above paper as J0409-56 E1, E2, E3 and E4. Rampazzo et al. (2020)
inspected also the corresponding areas  looking for \HII\ regions in
their frames. Those found are shown right panels of
Figure~\ref{NGC1533_Ha_outer} and indicated as E1+E2 and E4. Rampazzo et
al.  did not detect \HII\ regions in the zone indicated as E3. Indeed,
Werk et al. (2010) found that their targets span over a large range in
\Ha\ luminosities which correspond to a few O stars in  most of the
nearby cases. Werk et al. (2010) emphasized that often  FUV sources  
are mixed to  unresolved dwarf satellite companions and background
galaxies. Outer \HII\ regions  of NGC~1533 may be  linked to the strong
interacting phase  suggested by Cattapan et al. (2019) and well imaged
in their Figure~6. \HII\ regions detected and their embedded young stars
are definitely correlated with \HI\ as their velocities are the same
(see Section 6.2).

\section{Discussion}
\label{Discussion}

The NGC 1533 sub-structure in the Dorado group has been recently studied
by Cattapan et al. (2019) in $g$ and $r$ SDSS-bands and by Rampazzo et
al. (2020) in \HaN. This substructure with a heliocentric velocity
$V_{hel} \approx$ 764 \kms\ is still evolving separately from the Dorado
core, the compact group SCG 0414-5559 (Iovino 2002), with $V_{hel}
\approx$ 1230 \kms. Both IC 2038/ IC 2039 and NGC 1533 show several
interaction signatures as described in Cattapan et al. (2019 see their
Figure~6). Their  environment is \HI\ rich (Ryan-Weber et al. 2003,
Kilborn et al. 2009). The above studies suggest a common
evolutionary picture of the galaxies members of the NGC 1533
sub-structure.

\subsection{FUV and evolution of member galaxies}

Mazzei et al. (2014a, 2019 and references therein), using smoothed
particle hydrodynamic  simulations with chemo-photometric implementation
(SPH-CPI), investigated the evolutionary path of NGC 1533, the dominant
member of the sub-structure.

From a large grid of simulations of galaxy encounters and mergers,
starting from triaxial halos of gas and dark matter, a simulation
matching the global properties of this SB0 galaxy (absolute B mag, SED,
luminosity profile, morphology at different wavelengths and kinematics
properties) has been single out. According to this study NGC 1533 is the
result of a major merging occurred at $z=2.3$. 40\% of its  current mass
is assembled before $z=1$.

The FUV bright ring has been one of the morphological features used to
single out the simulation from the SPH-CPI grid. The ring is, indeed,
well reproduced by the selected simulation as the result of a resonance
that appeared when the galaxy is 8 Gyr old and is maintained up to now.
The simulation showed the path of NGC 1533 in the (NUV-r) vs. M$_r$
color magnitude diagram. NGC 1533 is 13.7 Gyr old, that it spends as
follows. It lies on the blue cloud for 7.2 Gyr, from there it takes
about 0.9 Gyr to reach the green valley that will cross reaching the red
sequence in 1.6 Gyr,  finally it gets its current position on the red
sequence after additional 4 Gyr.

The evolution of the pair IC 2038/ IC 2039 has not been studied yet. The
pair badly lack a detailed kinematic study in order to constrain the
SPH-CPI grid of simulations. As relevant examples in this context, we
mention the SPH-CPI studies of the pair NGC 454 by Plana et al. (2017)
and of the false pair NGC 3447/ NGC 3447A by Mazzei et al. (2018).

\subsection{FUV and SF regions}
\label{FUV_galaxies}

The FUV emission is a short scale ($\approx$ 10$^7$ years) SF indicator.
Therefore it can be associated to a measure of the SF obtained from \Ha\
($\approx$ 10$^6$ years)  (see Kennicutt \& Evans 2009). Next sections
will focus on SF regions and  SFR estimates from \Ha\ and FUV.

The FUV emission of IC 2038 is more extended than the area where \HII\
regions are found. Indeed, the FUV emission covers the entire galaxy
seen in the continuum (optical) image (Fugure~\ref{Tricromia_IC2038}
bottom panel). The FUV emission is also present in  the inner regions of
 IC 2039 (Figure~\ref{Tricromia_IC2038}). However, no \HII\ regions have
been detected in this E-S0 by Rampazzo et a. (2020).

At odds, in NGC 1533 \HII\ regions are found in two small complexes by
Rampazzo et al. (2020) and labelled as A and B in the  panel of
Figure~\ref{Tricromia_NGC1533}. In this galaxy the  FUV emission
concentrates in the ring and a bright FUV spot,  that includes the above
\HII\ regions.

\bigskip We revealed in this paper a faint FUV emission
(Figure~\ref{NGC1533_HI_UV}) associated to a complex \HI\
structures composed of two major arcs one north west and a second south
east  (Ryan-Weber et al. 2004). No optical counterparts are connected
with these arcs. Faint \HII\ regions are found south east of NGC
1533 (Werk et al. 2008, 2010; Rampazzo et al. 2020) and are correlated
to the \HI\ envelope. In particular regions indicated in
Figure~\ref{NGC1533_Ha_outer} as E1-E2 have a recession velocity of 831
and 846 \kms\ (another region indicated E5 by Ryan-Weber et al. (2004,
see their Figure ~2) has a recession velocity of 901 \kms). With these
recession velocities \HII\ regions are compatible with being associated
to NGC 1553 sub-structure (see Table~\ref{UVIT-sources}) and with the
\HI\ structure as well.

\medskip Concerning the origin of the \HI\ clouds around NGC 1533,
Ryan-Weber, Webster \& Bekki, K. (2003) suggested that it could be the
merger remnant of a tidally destroyed galaxy. Ryan-Weber et al.
(2004) noticed that the \HI\ gas in the south east cloud has velocity
dispersion up to 30 \kms\ and velocity gradient in the range 7– 50 \kms\
kpc$^{-1}$. These conditions  makes this site  unlikely  for SF since
the latter usually requires the gas to have a low velocity dispersion in
order to collapse.

 We conclude that the \HI\ arcs  detected by Ryan-Weber et al.
(2004), the correspondent faint FUV structure revealed in this paper in
addition to the optical evidence of faint tails and arcs shown  by the
Cattapan et al. (2019) deep surface photometry are indication that
galaxy-galaxy encounters, leading to galaxy tidal disruption (Ryan-Weber
et al. 2003) and merging events (Mazzei et al. 2019) are the drivers of
the complex evolution within the NGC 1533 sub-structure. 

\begin{figure}
	{\includegraphics[width=8.6cm]{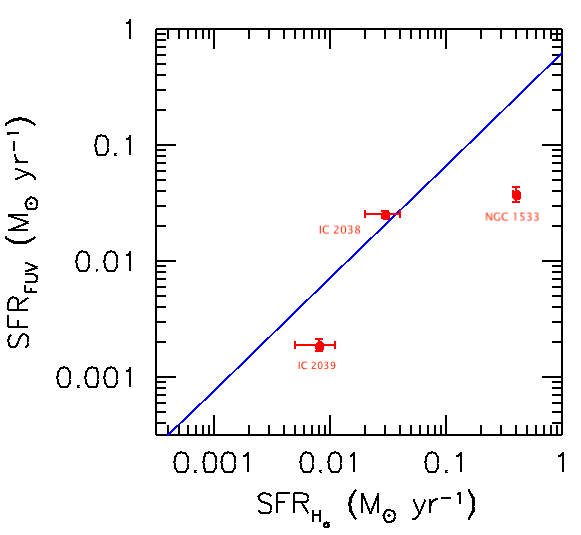}}
	
	\caption{Comparison between the SFR computed from H$_{\alpha}$  and FUV emission 
		without accounting for internal extinction.
		The blue solid line show the relation found by Lee et al. 2009 (see text).
	}	
	\label{SFR}
\end{figure}

\subsection{SFR from FUV integrated galaxy luminosity}
\label{SF_FUV}

In this section we compare the SFR for our three targets as derived by
\Ha\ emission (Rampazzo et al. 2020) with that by the integrated FUV
emission in this paper.

We   follow the recipes of Lee et al. (2009, their equation 3) to
compute the SFR for FUV:

$$	SFR~[M_\odot yr^{-1}] =1.4 \times 10^{-28} ~L_{FUV} ~[erg~s^{-1}
Hz^{-1}]~~ (1) $$

Lee et al. find the following relation when the effects of   internal
dust attenuation are not included:

$$ log (SFR(FUV)) = 0.79~log (SFR(H\alpha))-0.20~~(2) $$

This is highlighted by a solid blue line in Figure \ref{SFR}. Lee et al.
noticed that \Ha\ and FUV SFRs agree to within factor of $\approx$2 for
all galaxies with SFR $\geq 0.01$ M$_\odot$yr$^{-1}$.
 
The SFR of IC 2038, the only LTG in the sub-structure, do not deviate
from Lee et al. relatio However, according to Rampazzo et al. (2020) the
SFR of IC 2038 estimated from \Ha\ is below the average for its
morphological class estimated by James et al. (2004).

The SFR of the two ETGs, IC 2039 and NGC 1533 seems to deviate from the
Lee et al. relation, suggesting a higher SFR  from H$_{\alpha}$ than
from UV. This point is quite surprising given that UV stars trace SFR
with a time scale longer than H$_{\alpha}$ and the activity of SF is
residual  for ETGs.

Figure~2 and Table~2 in Lee et al. (2009) shows the trend of the ratio
log[SFR(\Ha)/SFR(FUV)] as a function of the B-band galaxy absolute
magnitude. At the distance of Dorado, the absolute B magnitude of NGC
1533 and IC 2039 is M$_B$=-19.52 and -16.3, respectively. At these
magnitudes, Lee et al. (2009) reported the average values of
log[SFR(\Ha)/SFR(FUV)]$\approx$-0.10$\pm$0.36 (1$\sigma$) and
log[SFR(\Ha)/SFR(FUV)]$\approx$-0.12$\pm$0.18. Our measured values are
log [SFR(\Ha)/SFR(FUV)]=1.02 and log [SFR(\Ha)/SFR(FUV)]=0.6 for NGC
1533 and IC~2039, respectively. However, Figure~2 (bottom panel) in Lee
et al. (2009) shows discrepant cases of galaxies, with values of log
[SFR(\Ha)/SFR(FUV)] similar to ours. Investigating the SF
properties in the Local Volume of Galaxies with \Ha\ and FUV fluxes,
Karachentsev \& Kaisina (2013) provided a log [SFR(\Ha)/SFR(FUV)]=1 for
NGC 1533, in well agreement with our value.

We plan to further explore the comparison of SFR from \Ha\ and FUV using
the entire {\tt UVIT} data-set of the Dorado backbone, considering
extinction effects in more detail than in Lee et al (2009) assumptions
(Rampazzo et al. in preparation). The SPH chemo-photometric simulation
of NGC 1533 by Mazzei et al. (2019) provided an estimate of the internal
galaxy extinction. We will use also such estimates.
 




\section{Summary and conclusion}
\label{Summary}

We performed with {\tt UVIT}  a FUV photometric study of a substructure of
the Dorado group of galaxies that includes three galaxies: the mixed 
morphology pair IC~2038 (Sbc)/2039 (E-S0) and NGC 1533.
We derived their luminosity profiles and discussed their FUV morphologies.

We found the following results:

\begin{itemize}

\item{The shape of the FUV luminosity profile indicates the presence of
a disc in all three galaxies. The presence of disc suggests that
dissipative mechanisms have been at work.}

\item{In IC 2038 the FUV emission is detected out to the optical size
for IC 2038, further out than the \HII\ regions system detected by
Rampazzo et al. (2020). There is no evidence of a XUV disk (Thilker
2008)}.

\item{In IC 2039 the FUV emission is detected in the inner regions where
no \HII\ regions have been detected by Rampazzo et al. (2020)}.

\item{In NGC 1533 the   FUV  emission is more extended than the system
of \HII\ regions detected by Rampazzo et al. (2020). The extended FUV
emitting regions likely correspond to outer arm-like structures detected
at different wavelengths} (Marino et al. 2011a, Rampazzo et al. 2017,
Cattapan et al. 2019).

\item{We reveal a faint FUV emission, just above the local background
noise, reminiscent of  the wide \HI\ structure detected by Ryan-Weber et
al. (2004). In the east and south east regions of this FUV emission lurk
 few \HII\ regions highlighted by   Werk at al. (2008,2010) and Rampazzo
et al. (2020) with the same redshift as the \HI\ structure (Ryan-Weber
at al. 2004) and of the NGC 1533 sub-structure as well.}

\item{We derive the SFR from the FUV luminosity and we compare the
results with SFR, following the Lee et al. (2009) recipe, with SFR from
\Ha\ by Rampazzo et al. (2020). The SFR  for IC 2038 only agrees with
Lee et al. (2009) expected ratio between FUV and \Ha\ derived values.
At odds, our measure of the SFR \Ha/FUV for NGC 1533 well agrees with
that found by Karachentsev \& Kaisina (2003).

Lee et al. (2009)   SFR(FUV) - SFR (Ha) relation does not account for
internal dust effects.  Such relation will be investigated further in
the analysis of our {\tt UVIT} FUV observations of the whole galaxy
group.}

\end{itemize}

In a forthcoming paper we will analyze the central part of Dorado,
the compact group SGC 0414-5559 Iovino (2002), composed of NGC 1553,
NGC 1549, NGC 1546 and IC 2058 plus the dwarf galaxy PGC 75125. The 
study of the SBc galaxy NGC 1536 and of the mixed pair NGC 1596/NGC 1602 
(S0+Irr), part of the A07 program but non yet observed, will complete our
study of the Dorado backbone with {\tt UVIT}.

\section*{Acknowledgements}

The {\tt UVIT} project is collaboration between the following institutes from India: 
Indian Institute of Astrophysics (IIA), Bengaluru, Inter University 
Centre for Astronomy and Astrophysics (IUCAA), Pune, and National 
Centre for Radioastrophysics (NCRA) (TIFR), Pune, and the 
Canadian Space Agency (CSA). The detector systems are provided by the 
Canadian Space Agency. The mirrors are provided by LEOS, ISRO, Bengaluru 
and the filter-wheels drives are provided by IISU, ISRO, Trivandrum. 
Many departments from ISAC, ISRO, Bengaluru have provided direct support in 
design and implementation of the various sub-systems.




\end{document}